\documentclass[pra,twocolumn,superscriptaddress,showpacs]{revtex4}
\usepackage{graphicx}
\usepackage{amsmath}
\usepackage{bbm}

\newcommand{\bra}[1]{\langle #1 |}
\newcommand{\ket}[1]{| #1 \rangle}
\newcommand{\braket}[2]{\langle #1 | #2 \rangle}
\newcommand{\Tr}{\hbox{Tr}}

\newcounter{wys}
\newcounter{odlwzgl}
\newcounter{pozstrzalki}

\newcommand{\spair}[4]{
\setcounter{wys}{#3}
\addtocounter{wys}{#4}
\setcounter{odlwzgl}{#2}
\addtocounter{odlwzgl}{-#1}
\put(#1,#3) {\circle*{5}}
\put(#1,#3) {\line(0,1){#4}}
\put(#1,\value{wys}) {\line(1,0){\value{odlwzgl}}}
\put(#2,#3) {\line(0,1){#4}}
\put(#2,#3) {\circle*{5}}
}

\begin{document}

\title{Immunity of information encoded in decoherence-free subspaces to particle loss}
\author{Piotr Migda{\l}}
\email{piotr.migdal@icfo.es}
\affiliation{Instytut Fizyki Teoretycznej, Wydzia{\l} Fizyki, Uniwersytet Warszawski, Ho\.{z}a 69, PL-00-681 Warszawa, Poland}
\affiliation{ICFO--Institut de Ci\`{e}ncies Fot\`{o}niques, 08860 Castelldefels (Barcelona), Spain}
\author{Konrad Banaszek}
\affiliation{Instytut Fizyki Teoretycznej, Wydzia{\l} Fizyki, Uniwersytet Warszawski, Ho\.{z}a 69, PL-00-681 Warszawa, Poland}
\pacs{03.67.Hk, 03.67.Pp, 42.50.Ex, 42.50.Dv}
\date{\today}

\begin{abstract}
We demonstrate that for an ensemble of qudits, subjected to collective decoherence in the form of perfectly correlated random SU($d$) unitaries, quantum superpositions stored in the decoherence free subspace are fully immune against the removal of one particle. This provides a feasible scheme to protect quantum information encoded in the polarization state of a sequence of photons against both collective depolarization and one photon loss, which can be demonstrated with photon quadruplets using currently available technology.
\end{abstract}

\maketitle

\section{Introduction}

Quantum systems are powerful yet fragile carriers of information. The ability to create and manipulate superposition states offers verifiably secure cryptography \cite{qcrypto}, reduces the complexity of certain computational problems \cite{qcomp}, and enables novel communication protocols \cite{qcomm}. However, in practical settings one needs to protect the quantum states carrying information against decoherence, i.e.\ uncontrolled interactions with the environment. This is accomplished by building redundancy into the physical implementation. Compared to the classical case, this task is much more challenging \cite{qec} due to limitations in handling quantum information, exemplified boldly by the no-cloning theorem \cite{nocloning}.

When an ensemble of elementary quantum systems decoheres through symmetric coupling with the environment, one can identify collective states that remain invariant in the course of evolution. These states span a so-called {\em decoherence-free subspace} (DFS) that is effectively decoupled from the interaction with the environment  \cite{dfs}. More generally, it is possible to identify subspaces that can be formally decomposed into a tensor product of two subsystems, one of which ``absorbs'' decoherence, while the second one, named  a {\em noiseless subsystem} or {\em a decoherence-free subsystem}, remains intact \cite{noiseless}.

In this paper we consider the DFS for an ensemble of $n$ qudits, i.e.\ elementary $d$-level systems, composed of states $\ket{\Psi}$ that are invariant with respect to an arbitrary perfectly correlated SU$(d)$ transformation:
\begin{equation}
\hat{U}^{\otimes n} \ket{\Psi} = \ket{\Psi}, \qquad \hat{U} \in \text{SU}(d).
\label{Eq:invariance}
\end{equation}
We show that this DFS features an additional degree of robustness, namely that the stored quantum information is immune to the loss of one of the qudits, regardless of the encoding. This result, specialized to the polarization state of single photons for which $d=2$, offers {\em combined} protection against two common optical decoherence mechanisms: photon loss \cite{photonloss} due to reflections, scattering, residual absorption, etc.\ as well as collective depolarization that occurs inevitably in optical fibers used for long-haul transmission \cite{birefringence,Bourennane2004}. Consequently, we provide here rigorous foundations to a speculation presented in Ref.~\cite{Boileau2004} that DFS-based quantum cryptography
can be made tolerant also to photon loss. It is worth noting that another physical realization of the qubit case can be also an ensemble of spin-$\frac{1}{2}$ particles \cite{Viola2001} coupled identically to a varying magnetic field.

The paper is organized as follows. First, in Sec.~\ref{Sec:SingletQubits} we briefly review the geometry of the singlet subspace for an ensemble of qubits and we explicitly show the robustness of the four qubit DFS, which spans the logical qubit space. This particular case leads us to a proposal for a proof-of-principle experiment based on currently available photonic technologies that demonstrates the robustness of DFS encoding, presented in Sec.~\ref{Sec:ExperimentalScheme}. The general proof for an arbitrary $d$ that a quantum superposition encoded in an $\text{SU}(d)$ DFS remains immune against the loss of one particle is described in Sec.~\ref{Sec:General}. Finally, Sec.~\ref{Sec:Conclusions} concludes the paper.

\section{Singlet subspace for qubits}
\label{Sec:SingletQubits}

Because of two relevant physical realizations using photons and spin-$1/2$ particles, we will first discuss the qubit case with $d=2$. The complete Hilbert space of an ensemble of $n$ qubits, each described by a two-dimensional spin-$1/2$ space $\mathcal{H}_{1/2}$, can be subjected to Clebsch-Gordan decomposition \cite{Dicke1954}
\begin{equation}
(  \mathcal{H}_{1/2} )^{\otimes n} = \bigoplus_{j = (n \bmod 2)/2 }^{n/2} \mathbbm{C}^{K^j_n} \otimes \mathcal{H}_{j}\label{eq:clebsch},
\end{equation}
where the direct sum is taken with the step of one and
$K^j_n$ are multiplicities of spin-$j$ Hilbert spaces $\mathcal{H}_{j}$, given explicitly by
\begin{equation}
K^j_n  = \frac{2j+1}{n/2+j+1} {n \choose n/2+j}\label{eq:catalan}.
\end{equation}
The action of $\hat{U}^{\otimes n}$, where $\hat{U}$ is any $\text{SU}(2)$ transformation, affects only $\mathcal{H}_{j}$ in Eq.~(\ref{eq:clebsch}), leaving $\mathbbm{C}^{K^j_n}$ unchanged. In particular, for an even number of $n$ qubits forming the ensemble, the {\em singlet subspace} corresponding to $j=0$ is free from decoherence. Furthermore, removing one particle from that ensemble maps any initial state from the singlet subspace onto a certain state from the {\em doublet subspace} $\mathbbm{C}^{K^{1/2}_{n-1}} \otimes \mathcal{H}_{1/2}$. Because $K^{1/2}_{n-1}=K^0_n$, it is plausible that the quantum superposition will end up entirely in the decoherence-free subsystem $\mathbbm{C}^{K^{1/2}_{n-1}}$ where it will remain protected from collective depolarization.

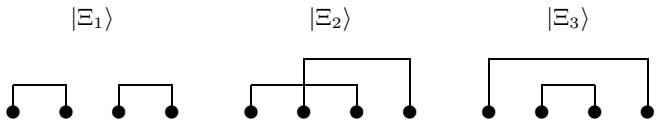
\begin{figure}
\centering
\begin{picture}(240,45)

\put(30,40) {\makebox(0,0)[c]{$\ket{\Xi_1}$}}
\spair{0}{20}{5}{10}
\spair{40}{60}{5}{10}

\put(120,40) {\makebox(0,0)[c]{$\ket{\Xi_2}$}}
\spair{90}{130}{5}{10}
\spair{110}{150}{5}{20}

\put(210,40) {\makebox(0,0)[c]{$\ket{\Xi_3}$}}
\spair{180}{240}{5}{20}
\spair{200}{220}{5}{10}

\end{picture}
\caption{Diagrams depicting three non-equivalent products of two-qubit singlet states defined in Eq.~(\protect\ref{eq:dfs4}). The qubits are represented as dots with connections identifying pairs that form singlet states.}
\label{fig:spps}
\end{figure}

The simplest non-trivial case is $n=4$ physical qubits encoding one logical qubit. Let us consider three states from the four-qubit DFS defined as products
\begin{align}
\ket{\Xi_1} &= \ket{\psi^-}_{12} \ket{\psi^-}_{34}, \nonumber\\
\ket{\Xi_2} &= \ket{\psi^-}_{13} \ket{\psi^-}_{42}, \label{eq:dfs4}\\
\ket{\Xi_3} &= \ket{\psi^-}_{14} \ket{\psi^-}_{23}, \nonumber
\end{align}
where $\ket{\psi^-}_{ij}= (\ket{01}_{ij}-\ket{10}_{ij})/\sqrt{2}$ is the singlet state of qubits $i$ and $j$. These states, shown schematically in Fig.~\ref{fig:spps}, form an overcomplete set in the DFS. For concreteness, let us select $\ket{\Xi_1}$ and $\ket{\Xi_3}$ as a non-orthogonal basis.
Any state of the logical DFS qubit can be written as a superposition
\begin{align}
\ket{\Psi} = \alpha \ket{\Xi_1}+ \beta \ket{\Xi_3},
\label{Eq:Psifourqubit}
\end{align}
where $\alpha$ and $\beta$ are complex amplitudes. Without loss of generality we can assume that the first physical qubit has been lost. The remaining three qubits are described by an equally weighted statistical mixture of two states:
\begin{align}
\ket{\Psi^{(0)}}_{\bar{1}} &= \alpha \ket{1}_{2} \ket{\psi^-}_{34} + \beta  \ket{\psi^-}_{23} \ket{1}_{4} \nonumber \\
\ket{\Psi^{(1)}}_{\bar{1}} &= \alpha \ket{0}_{2} \ket{\psi^-}_{34} + \beta  \ket{\psi^-}_{23} \ket{0}_{4},\label{eq:singletowyrozpisany}
\end{align}
where $\ket{\cdot}_{\bar{1}}$ denotes the state of all qubits but the first one. It is easy to see that a collective transformation
$\hat{U}^{\otimes 3}$ leaves the statistical mixture $\frac{1}{2} \bigl( \ket{\Psi^{(0)}}_{\bar{1}} \bra{\Psi^{(0)}} +
\ket{\Psi^{(1)}}_{\bar{1}} \bra{\Psi^{(1)}}\bigr)$ intact.

After the loss of the first particle, the initial four-qubit state from Eq.~(\ref{Eq:Psifourqubit}) can be recovered through the following procedure.
First, one needs to measure in a non-destructive way the $z$ component of the total pseudospin operator $\hat{\sigma}^{z}_{2} + \hat{\sigma}^{z}_{3} + \hat{\sigma}^{z}_{4}$, where $\hat{\sigma}^{z} = \ket{0}\bra{0} - \ket{1}\bra{1}$, in order to discriminate $\ket{\Psi^{(0)}}$ from $\ket{\Psi^{(1)}}$. If the result corresponding to $\ket{\Psi^{(1)}}$ is obtained, we apply a collective rotation $(\hat{\sigma}^{x})^{\otimes 3}$, where $\hat{\sigma}^x = \ket{0}\bra{1} + \ket{1}\bra{0}$. This yields the state $\ket{\Psi^{(0)}}_{\bar{1}}$. In the second step, one replaces the lost qubit with a new one prepared in a state $\ket{+}_{1} = \frac{1}{\sqrt{2}}( \ket{0}_1 + \ket{1}_1 )$ and applies a controlled rotation which restores the original state $\ket{\Psi}$:
\begin{equation}
\bigl( \ket{0}_{1} \bra{0} \otimes {\hat{\openone}}^{\otimes 3} + \ket{1}_{1} \bra{1} \otimes {(\hat{\sigma}^x)}^{\otimes 3}\bigr)
\bigl( \ket{+}_{1} \ket{\Psi^{(0)}}_{\bar{1}}\bigr)  = \ket{\Psi}
\end{equation}
Note that this rotation can be realized as a sequence of three C-NOT gates.

The robustness of DFS to particle loss can be intuitively understood in the following way. DFS states owe their invariance with respect to collective unitary transformation to a very rigid structure. In fact, if we write  a DFS state as a superposition in the computational basis for individual qubits, the state of one qubit can be determined unambiguously from the states of the remaining ones. This suggests that the loss of one particle does not destroy any information. Futher, it is always possible to repair the state as there is only one unique way to fit the lost particle such that the singlet symmetry is recovered.

\section{Experimental scheme}
\label{Sec:ExperimentalScheme}

We will now present a proposal a feasible experiment that demonstrates the robustness of DFS encoding using
photon quadruplets that can be generated in the process of parametric down-conversion \cite{Bourennane2004,Weinfurter2001Gong2008}.
The basis states $\ket{0}$ and $\ket{1}$ correspond in this case to horizontal and vertical polarizations of individual photons.
Let us consider four-photon states $\ket{\Xi_k}$, $k=1,2,3$, defined in Eq.~(\ref{eq:dfs4}) as well as their orthogonal complements in the two-dimensional DFS, which we will denote as $\ket{\Xi_k^\perp}$.
The index $k$ corresponds to three non-equivalent orderings of the photons and it can be changed by suitable rerouting of the photons. As demonstrated in \cite{Bourennane2004}, the states $\ket{\Xi_1}$ and $\ket{\Xi_1^\perp}$ can be discriminated unambiguously by detecting polarizations in the horizontal-vertical basis $\ket{0}, \ket{1}$ for photons $12$ and in the diagonal basis $(\ket{0} \pm \ket{1})/\sqrt{2}$ for photons $34$. Restricted to the DFS subspace, this strategy yields the standard projective measurement.

\begin{figure}
\centering
\begin{tabular}{rl}
\includegraphics[width=0.22\textwidth]{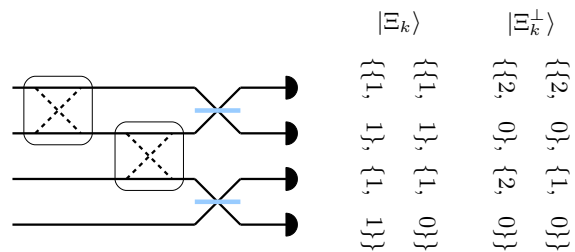} & 	
	\begin{picture}(100,70)
\put(30,85) {\makebox(0,0)[c]{$\ket{\Xi_k}$}}
\put(20,35){\makebox(0,0)[c]{\rotatebox{270}{$\{ \{ 1,\ \ 1 \},\ \{1,\ \ 1\} \}$}}}
\put(40,35){\makebox(0,0)[c]{\rotatebox{270}{$\{ \{ 1,\ \ 1 \},\ \{1,\ \ 0\} \}$}}}
\put(80,85) {\makebox(0,0)[c]{$\ket{\Xi_k^\perp}$}}
\put(70,35){\makebox(0,0)[c]{\rotatebox{270}{$\{ \{ 2,\ \ 0 \},\ \{2,\ \ 0\} \}$}}}
\put(90,35){\makebox(0,0)[c]{\rotatebox{270}{$\{ \{ 2,\ \ 0 \},\ \{1,\ \ 0\} \}$}}}
	\end{picture}
\end{tabular}
\caption{An experimental scheme for loss-tolerant detection of a logical qubit encoded in four photons. The projection basis $\ket{\Xi_k}$, $\ket{\Xi_k^\perp}$, where $k=1,2,3$, is selected by a suitable rerouting of input photons. Pairs of photons are interfered on two balanced beam splitters and photon numbers are counted at their outputs. Combinations of outcomes for individual detectors that correspond to unambiguous identification of $\ket{\Xi_k}$ and $\ket{\Xi_k^\perp}$ are indicated with photon numbers in curly brackets. The ordering within both inner and outer brackets does not matter.}
	\label{fig:measurement}
\end{figure}

It is easy to check that the above individual measurement no longer works if one of the photons is missing. It turns out that this problem can be solved by resorting to collective measurements. Suppose that we interfere photon pairs $12$ and $34$ on two separate balanced beam splitters, playing the role linear-optics Bell state analyzers \cite{BrauMannPRA94}. The state $\ket{\Xi_1}$ will yield exactly one photon in each output port of each beam splitter. In contrast, because the orthogonal state $\ket{\Xi_1^\perp}$ can be written as \cite{noiseless}:
\begin{equation}
\ket{\Xi_1^\perp} = \frac{1}{\sqrt{3}}\left(
\ket{00}_{12}\ket{11}_{34}
+ \ket{11}_{12}\ket{00}_{34} - \ket{\psi^+}_{12}\ket{\psi^+}_{34} \right),
\end{equation}
where $\ket{\psi^+}_{ij}= (\ket{01}_{ij}+\ket{10}_{ij})/\sqrt{2}$, it will always produce two photons at the same output port for each of the two beam splitters. If one photon is lost, the states $\ket{\Xi_1}$ and $\ket{\Xi_1^\perp}$ will still give distinguishable outcomes: registering two photons at a single output unambiguously heralds $\ket{\Xi_1^\perp}$, while registering a photon pair at two different outputs of the same beam splitters detects $\ket{\Xi_1}$. The third photon will emerge separately from the second beam splitter. This detection scheme is summarized in Fig.~\ref{fig:measurement}.

An interesting question is whether the scheme described above could be exploited for quantum key distribution.
The scalar products between any two the states $\ket{\Xi_k}$ and $\ket{\Xi_l}$ with $k \neq l$ are equal to $\braket{\Xi_k}{\Xi_l} = -\frac{1}{2}$. In the Bloch representation of the two-dimensional DFS, they form a regular triangle inscribed into a great circle on the Bloch sphere,
constituting a so-called {\em trine} that warrants cryptographic security \cite{Boileau2004,Renes2004Tabia2011}. To generate a key, the sender Alice could prepare photon quadruplets in one of randomly selected states $\ket{\Xi_1}$, $\ket{\Xi_2}$, or $\ket{\Xi_3}$. The ability to perform a projection onto any pair of orthogonal states $\ket{\Xi_k}, \ket{\Xi_k^\perp}$ would enable the receiving party Bob to tell, in the case when an outcome $\ket{\Xi_k^\perp}$ is obtained, which state has definitely not been prepared by Alice. Such correlations between Alice's preparations and Bob's outcomes can be distilled into a secure key.

We have shown that the projective measurement onto $\ket{\Xi_k}, \ket{\Xi_k^\perp}$ can be implemented in a way that tolerates the loss of one photon. In a cryptographic setting, the crucial issue is to ensure that an eavesdropper Eve does does not map the state of intercepted photons outside the DFS, which may enable eavesdropping attacks beyond those already studied  \cite{Boileau2004,Renes2004Tabia2011}. To verify that this is not the case, Bob could perform in principle a full quantum state reconstruction on some of the transmissions, which however would be resource consuming. We conjecture that a sufficient strategy to detect such an attack would be: (i) to detect polarizations of photons emerging after the beam splitters; (ii) for a subset of transmissions to count directly received photons to ensure that no multiphoton states in individual input paths occur; (iii) for another subset of transmissions to apply before the beam splitters random and uncorrelated transformations $\hat{U} \otimes \hat{U}$ and $\hat{U}' \otimes \hat{U}'$ and check that states $\ket{\Xi_k}$ always yield the correct outcome when Bob used the matching basis for his measurement.

\section{General proof}
\label{Sec:General}

The reasoning presented in Sec.~\ref{Sec:SingletQubits} can be generalized to any even number of $n > 4$ qubits by considering DFS states given by products of two-qubit singlet states. Such states form an overcomplete set in the DFS \cite{Lyons2008}, which enables one to follow directly the steps described for four qubits. The robustness of DFS encoding can be shown more generally for an ensemble of $n$ qudits, i.e.\ $d$-dimensional systems.
In this case, a DFS satisfying Eq.~\eqref{Eq:invariance}  exists only when $n$ is a multiple of $d$, which follows from the structure of the Young tableux for irreducible representations of tensor products of the $SU(d)$ group \cite{Jones1998groupsrepresentations}.

As before, for concreteness we will consider removal of the first qudit. Let us consider arbitrary two states $\ket{\Psi}$ and $\ket{\Phi}$ from the DFS and expand them in the form analogous to Eq.~(\ref{eq:singletowyrozpisany}):
\begin{equation}
\ket{\Psi} = \frac{1}{\sqrt{d}} \sum_{i=0}^{d-1} \ket{i}_1 \ket{\Psi^{(i)}}_{\bar{1}},
\quad
\ket{\Phi} = \frac{1}{\sqrt{d}} \sum_{i=0}^{d-1} \ket{i}_1 \ket{\Phi^{(i)}}_{\bar{1}}
\label{eq:sud_singletowyrozpisany}
\end{equation}
where $\ket{i}_1 , i=0,\ldots,d-1$ is an orthonormal basis in the space of the first qudit, and
$\ket{\Psi^{(i)}}_{\bar{1}} = \sqrt{d}\, {}_{1}\! \braket{i}{\Psi}$ and $\ket{\Phi^{(i)}}_{\bar{1}} = \sqrt{d} \, {}_{1}\! \braket{i}{\Phi}$ are states of the remaining $n-1$ qudits. We will first show that the following  general property holds:
\begin{equation}
{}_{\bar{1}} \! \braket{\Phi^{(i)}}{\Psi^{(j)}}_{\bar{1}} = \delta_{ij} \braket{\Phi}{\Psi}.
\label{Eq:SUd:dmuidmuj}
\end{equation}
As we will see, this property guarantees that the loss of one particle does not destroy the quantum information encoded in the DFS.

In order to show that for $i \neq j$ the states $\ket{\Phi^{(i)}}$ and $\ket{\Psi^{(j)}}$  are orthogonal as implied by Eq.~\eqref{Eq:SUd:dmuidmuj}, let us consider the action of
a diagonal unitary operator $\hat{V}^{\otimes n}$, where $\hat{V} = \text{diag} (e^{i \phi_0}, \ldots, e^{i \phi_{d-1}} )$ with arbitrary phases $\phi_0, \ldots, \phi_{d-1}$ that sum up to zero. Invariance of $\ket{\Phi^{(i)}}_{\bar{1}}$ and $\ket{\Psi^{(j)}}_{\bar{1}}$ under $\hat{V}^{\otimes n}$ implies that in the basis formed by tensor products of states $\ket{0}, \cdots, \ket{d-1}$ they are composed only from terms that have exactly $n/d$ particles in each of these $d$ states. Consequently, projecting the first qudit on orthogonal states $\ket{i}_{1}$ and $\ket{j}_{1}$ leaves the remaining qudits in distinguishable states.

In order to verify the case when $i=j$ in Eq.~(\ref{Eq:SUd:dmuidmuj}) it is convenient to use the transformation of states $\ket{\Psi^{(i)}}_{\bar{1}}$ under the action of $\hat{U}^{\otimes (n-1)}$. In order to derive this transformation, let us rewrite the invariance condition from Eq.~\eqref{Eq:invariance} to the form
$\hat{U}^\dagger \otimes \openone^{\otimes (n-1)} \ket{\Psi} = \openone \otimes \hat{U}^{\otimes (n-1)} \ket{\Psi}$ and  project the first qudit onto $ \sqrt{d} \, {}_{1} \! \bra{i}$. This yields the identity:
\begin{equation}
\hat{U}^{\otimes (n-1)} \ket{\Psi^{(i)}}_{\bar{1}} =
\sqrt{d} \, \bigl( {}_{1} \! \bra{i} \hat{U}^\dagger \bigr) \ket{\Psi}
=
\sum_{j=0}^{d-1} \bigl( \bra{j} \hat{U} \ket{i} \bigr)^\ast \ket{\Psi^{(j)}}_{\bar{1}}
\label{Eq:iUdagger}
\end{equation}
Let us now specialize this result to a special unitary transformation that cyclically shifts the labelling of the basis states:
\begin{equation}
\hat{W} = (-1)^{d-1} \sum_{i=0}^{d-1}  \ket{i+ 1 }\bra{i},\label{eq:unitaryW}
\end{equation}
where the addition $i+1$ is understood to be modulo $d$. Using this $\hat{W}$ in Eq.~\eqref{Eq:iUdagger} implies that $\ket{\Psi^{(i+1)}} =(-1)^{d-1} \hat{W}^{\otimes (n-1)} \ket{\Psi^{(i)}}$, i.e.\ $\ket{\Psi^{(i)}}$ and $\ket{\Psi^{(i+1)}}$ are related by a unitary that is independent of $\ket{\Psi}$. This means that
$\braket{\Phi^{(i+1)}}{\Psi^{(i+1)}} = \braket{\Phi^{(i)}}{\Psi^{(i)}}$. This fact combined with expanding the scalar product $\braket{\Phi}{\Psi}$ using Eq.~(\ref{eq:sud_singletowyrozpisany}) completes the proof of  Eq.~(\ref{Eq:SUd:dmuidmuj}).

With Eq.~(\ref{Eq:SUd:dmuidmuj}) in hand, further steps are straightforward.
A removal of the first qudit maps a state $\ket{\Psi}$ onto a statistical mixture
\begin{equation}
\hat{\varrho}_{\bar{1}}  = \Tr_1 \bigl( \ket{\Psi}\bra{\Psi} \bigr) =\frac{1}{d} \sum_{i=0}^{d-1} \ket{\Psi^{(i)}}_{\bar{1}}\bra{\Psi^{(i)}}.
\end{equation}
Eq.~(\ref{Eq:SUd:dmuidmuj}) implies that analogously to the SU(2) case the components with different $i$ occupy orthogonal subspaces. Within each subspace the state is fully preserved, which follows from applying Eq.~(\ref{Eq:SUd:dmuidmuj}) to pairs of states from an arbitrary basis in the DFS. The final step is to show that the state $\hat\varrho_{\bar{1}}$ is invariant with respect to $\hat{U}^{\otimes (n-1)}$. This is a consequence of the fact that both the initial state $\ket{\Psi}$ and the procedure of tracing out a particle are invariant with respect to SU($d$) transformations. Explicitly, the invariance of $\hat\varrho_{\bar{1}}$ can be verified with a calculation based on Eq.~(\ref{Eq:iUdagger}):
\begin{multline}
\hat{U}^{\otimes (n-1)} \hat{\varrho}_{\bar{1}} (\hat{U}^\dagger)^{\otimes (n-1)} =
\sum_{i=0}^{d-1} \bigl( {}_{1} \! \bra{i} \hat{U}^\dagger \bigr) \ket{\Psi}\bra{\Psi} \bigl( \hat{U} \ket{i}_{1} \bigr) \\
= \Tr_1 \bigl( \ket{\Psi}\bra{\Psi} \bigr) = \hat{\varrho}_{\bar{1}}.
\end{multline}
Thus the encoded state is fully preserved.

\section{Conclusions}
\label{Sec:Conclusions}

Concluding, we have shown that DFS encoding is immune to removing one particle. Unfortunately, this property does not seem to generalize in a straightforward manner to the loss of more particles. For example, when two qubits are removed from a four-qubit DFS state, the result will be either a singlet state of the remaining two qubits, or a statistical mixture of the singlet and triplet states which does not preserve the original superposition. This observation holds also for any higher even number of qubits. Nevertheless, our result shows how to protect information in the few-photon regime from both collective depolarization and the first-order effects of linear attenuation. We have proposed an experimental demonstration of this combined protection which can provide a robust quantum cryptography protocol.

Finally, let us note that although the proof of robustness against the qudit loss was based on the assumption that Eq.~(\ref{Eq:invariance}) is satisfied for every $\text{SU}(d)$ matrix, the DFS fulfilling this condition protects quantum superpositions from any decoherence mechanism that involves a subset of $\text{SU}(d)$ transformations. Therefore our considerations apply to a range of physical systems, for example higher-spin particles in a magnetic field or multilevel atoms interacting with optical fields.

We acknowledge useful discussions with Rafa{\l} Demkowicz-Dobrza\'{n}ski, Harald Weinfurter, Alessio Celi, Marcin Kotowski and Joanna Zieli\'{n}ska. The work was supported by the Foundation for Polish Science TEAM project cofinanced by the EU European Regional
Development Fund, FP7 FET project CORNER (contract no. 213681), Spanish MINCIN project FIS2008-00784 (TOQATA) and ERC Advanced Grant QUAGATU.


\begin{thebibliography}{20}

\bibitem{qcrypto}
N. Gisin, G. Ribordy, W. Tittel, and H. Zbinden,
Rev. Mod. Phys. {\bf 74}, 145 (2002) arXiv:quant-ph/0101098v2;
V. Scarani, H. Bechmann-Pasquinucci, N. J. Cerf, M. Du\v{s}ek, N. L\"{u}tkenhaus, and M. Peev, {\em ibid.} {\bf 81}, 1301 (2009) arXiv:0802.4155v3.

\bibitem{qcomp}
A. Ekert and R. Jozsa, Rev. Mod. Phys. {\bf 68}, 733 (1996).

\bibitem{qcomm}
H. Buhrman, R. Cleve, S. Massar, and R. de Wolf,
Rev. Mod. Phys. {\bf 82}, 665 (2010).

\bibitem{qec}
D. Gottesman, in {\em Proceedings of Symposia in Applied Mathematics}, {\bf 68}, pp. 13-58 (2010).

\bibitem{nocloning}
W.K. Wootters and W.H. Zurek,  Nature {\bf 299}, 802 (1982).

\bibitem{dfs}
L.-M. Duan and G.-C. Guo, Phys. Rev. Lett. {\bf 79}, 1953 (1997) arXiv:quant-ph/9703040v2;
P.~Zanardi and M.~Rasetti, {\it ibid.} {\bf 79}, 3306 (1997) arXiv:quant-ph/9705044v2;
D. A. Lidar, I. L. Chuang, and K. B. Whaley, {\it ibid.} {\bf 81}, 2594 (1998) arXiv:quant-ph/9807004v2.

\bibitem{noiseless}
J.~Kempe, D.~Bacon, D.~Lidar, and K.~B.~Whaley,
Phys. Rev. A {\bf 63}, 042307 (2001) arXiv:quant-ph/0004064v2.

\bibitem{photonloss}
W. Wasilewski and K. Banaszek, Phys. Rev. A {\bf 75}, 042316 (2007) arXiv:quant-ph/0702075v1;
C.-Y. Lu, W.-B. Gao, J. Zhang, X.-Q. Zhou, T. Yang, and J.-W. Pan,
Proc. Natl Acad. Sci. USA {\bf 105}, 11050 (2008).

\bibitem{birefringence}
S. D. Bartlett, T. Rudolph, and R.W. Spekkens, Phys. Rev.
Lett. {\bf 91}, 027901 (2003) arXiv:quant-ph/0302111v3;
K. Banaszek, A. Dragan, W. Wasilewski, and C. Radzewicz,
{\em ibid.} {\bf 92}, 257901 (2004) arXiv:quant-ph/0403024v1.


\bibitem{Bourennane2004}
M.~Bourennane, M.~Eibl, S.~Gaertner, C.~Kurtsiefer, A.~Cabello, and
  H.~Weinfurter,
Phys. Rev. Lett. {\bf 92}, 107901 (2004) arXiv:quant-ph/0309041v2.

\bibitem{Boileau2004}
J.~C. Boileau, D.~Gottesman, R.~Laflamme, D.~Poulin, and R.~W. Spekkens,
\newblock {Phys. Rev. Lett.} {\bf 92}, 017901 (2004) arXiv:quant-ph/0306199v2.

\bibitem{Viola2001}
L.~Viola, E.~M.~Fortunato, M.~A.~Pravia, E.~Knill, R.~Laflamme, and D.~G.~Cory,
Science {\bf 293}, 2059 (2001).

\bibitem{Dicke1954}
R.~H.~Dicke, Phys. Rev. {\bf 93}, 99 (1954).

\bibitem{Weinfurter2001Gong2008}
H.~Weinfurter and M.~\.{Z}ukowski,
\newblock {Phys. Rev. A} {\bf 64}, 010102 (2001) arXiv:quant-ph/0103049v1;
Y.-X. Gong, X.-B. Zou, X.-L. Niu, J.~Li, Y.-F. Huang, and G.-C. Guo,
\newblock {Phys. Rev. A} {\bf 77}, 042317 (2008).

\bibitem{BrauMannPRA94}
S. L. Braunstein and A. Mann, Phys. Rev. A {\bf 51}, R1727 (1995).

\bibitem{Renes2004Tabia2011}
J.~M.~Renes,
\newblock {Phys. Rev. A} {\bf 70}, 052314 (2004) arXiv:quant-ph/0402135v3;
G.~Tabia and B.-G. Englert,
\newblock {Phys. Lett. A}, {\bf 375}, 817 (2011) arXiv:0910.5375v1.

\bibitem{Lyons2008}
D.~W. Lyons and S.~N. Walck,
\newblock {Phys. Rev. A} {\bf 78}, 042314, (2008) arXiv:0808.2989v2.

\bibitem{Jones1998groupsrepresentations}
H.~F. Jones,
\newblock {\em {Groups, Representations and Physics}}.
\newblock (Institute of Physics Publishing, Bristol and Philadelphia, 1998), Chap. 8.



\end{thebibliography}
\end{document}